# Optical Mixing Controlled Stimulated Scattering instabilities: Suppression of SRS by the Controlled Introduction of Ion Acoustic and Electron Plasma Wave Turbulence

Bedros Afeyan,[a] M. Mardirian,[a] K. Won,[a] D. S. Montgomery,[b] J. Hammer,[c] R. K. Kirkwood,[c] A. J. Schmitt[d]

[a] Polymath Research Inc., Pleasanton, CA
[b] Los Alamos National Laboratory, Los Alamos, NM
[c] Lawrence Livermore National Laboratory, Livermore, CA
[d] Naval Research Laboratories, Washington, DC

**ABSTRACT**

In a series of experiments on the Omega laser facility at LLE, we have demonstrated the suppression of SRS in prescribed spectral windows due to the presence of externally controlled levels of ion acoustic waves (IAW, by crossing two blue beams at the Mach -1 surface) and electron plasma waves (EPW, by crossing a blue and a green beam around a tenth critical density plasma) generated via optical mixing. We have further observed SRS backscattering of a green beam when crossed with a blue pump beam, in whose absence, that (green beam) backscattering signature was five times smaller. This is direct evidence for green beam amplification when crossed with the blue. Additional proof comes from transmitted green beam measurements. A combination of these techniques may allow the suppression of unacceptable levels of SRS near the light entrance hole of large-scale hohlraums on the NIF or LMJ.

## 1. INTRODUCTION

The success of laser based inertial confinement fusion (ICF) depends critically on the ability to convert the incident laser energy to X rays at the walls of a hohlraum, in the case of indirect drive,[1] and to plasma thermal energy so as to drive an ablation front, in the case of direct drive.[2] Parametric instabilities[3] inevitably occur when high enough intensity ($I_{laser} \sim 10^{15}$ W/cm$^2$) coherent laser radiation is introduced into a low enough temperature ($T_e < 5$ keV), and high enough density ($n_e/n_c > .05$) and long enough scale length ($L_n >$ 100s of μms) plasmas. Such coherent wave-wave interactions,[3] for instance, SRS, SBS and filamentation, must be mitigated so that an inadmissibly high fraction of the incident laser energy is not misdirected via nonlinear scattering losses. For the success of ICF, excessive hot electron preheat can not be tolerated either, but which may, nevertheless, occur with SRS and $2\omega_{pe}$.[1-3]

For the last four years, on Omega, at LLE, as part of the NLUF program, we have studied ways to mitigate the severity of these processes. Our idea was to use crossing laser beams in order to introduce controlled levels of plasma fluctuations and turbulence so as to make the plasma inhospitable to coherent, highly correlated three wave interactions. We have already shown[4] that at the Mach −1 surface, when two blue beams cross and generate a large amplitude IAW, they strongly diminish the SRBS of the blue beam coming from the vicinity of that density. This is shown in Fig. 1 both for weak and strong IAW damping cases. We can see from Fig. 1a that the reduction of the reflectivity of SRS (in the proper wavelength window corresponding to the vicinity of the Mach −1 surface) in the presence of a large amplitude IAW (driven by two high intensity lasers) is by a factor of 3-5 when compared to the SRS levels generated when the beam crossing the pump beam is much weaker (1/15$^{th}$ the intensity of the pump) and hence unable to create large amplitude IAWs. Even more so, in the case of weak IAW damping, (where Be targets were used which do not have Hydrogen in them as do the CH targets) the effect was a factor of 7 to 8 SRS reflectivity reduction from the proper wavelength window when comparing 1:1 to 1/10:1 intensity ratio cases between probe and pump laser beams, respectively. Fig. 2 shows SRS suppression by a factor 2-3 when large amplitude EPWs are optically generated and act as hindrances to a witness beam's SRBS process.

More recently, we have conducted experiments where we crossed high intensity blue and green beams that could, around a tenth critical density (for the blue), resonantly drive an EPW. Fig. 3 shows the experimental configuration for these experiments. We have various means of diagnosing the resonant interactions between crossing blue and green beams. One novel way is via the observation of green beam SRBS with spectral signatures far weaker when the crossing blue beam is not present. We interpret this new result as the amplification of the Green beam via optical mixing with the blue pump beam and the subsequent Raman backscattering of the green beam from the peak of the exploding foil density profile when its intensity is boosted so as to become high enough to produce significant levels of SRBS. This is quite interesting since it shows that energy transfer between crossing beams leaves the amplified beam coherent enough to be able to instigate high levels of backscattering of its own.

In our next set of experiments, we hope to study these phenomena more carefully and to combine these processes to see just how far we can go to suppress parametric instabilities in fluctuating and turbulent plasmas without inadvertently triggering some other undesirable process.

## 2. EXPERIMENTAL RESULTS

We have shot 7-10 µm CH exploding foils as well as 5 µm Be foils to create sub-tenth-critical plasmas with parabolic density profiles. In our most recent experiments, with crossing blue and green laser beams, four 500 J beams per side were used for 1 ns to create the exploding foil plasma and four more per side to heat it for another ns. In the middle of the second set of heater beams, the interaction beams were introduced.

Fig. 2 shows the suppression of SRS of a witness beam (BL30) in the presence of crossing blue (BL65) and green beams, which resonantly generate an EPW as the appropriate (roughly a ninth critical density for the blue) density is reached at the peak of the exploding foil plasma.

Fig. 3 shows the configuration of the experiment when BL60 as opposed to 65 was used as the blue pump beam. The Green beam TBD introduced an occlusion in front of BL65. We observed enhanced green beam amplification due to the presence of the blue pump beam. This is ascertained in Fig. 4 by the novel and unexpected measurement of significant green beam Raman backscatter enhancement (by a factor of 15) triggered directly by the green beam being crossed with a blue beam.

## ACKNOWLEDGMENTS

This work was supported by DOE Grants DE-FG03-03SF22690 and DE-FG03-03NA00059. We would like to thank the Omega crew at LLE as well as J. Soures, R. Bahr, S. Morse, G. Pien and M. Bonino, for their invaluable support and O. Landen and LLNL for providing a wider spot DPP and a TBD diagnostic for the green beam which made the new results reported here possible.

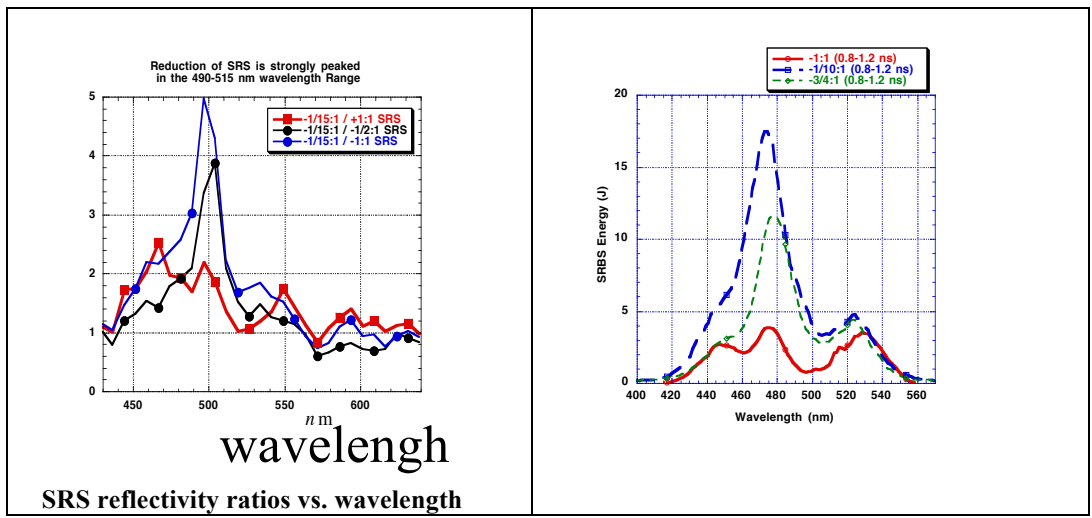

SRS reflectivity ratios vs. wavelength

**Figure 1.** Reduction of Raman backscattering levels in the presence of two high intensity crossed blue beams compared to scattering levels when the probe beam is weak in a) 10 μm CH targets which correspond to high IAW damping (reduction is a factor of 5 in the wavelength window around the Mach –1 surface where resonant interaction is possible) and b) 5 μm Be targets which correspond to the weak IAW damping regime with equivalent hydrodynamic expansion properties.

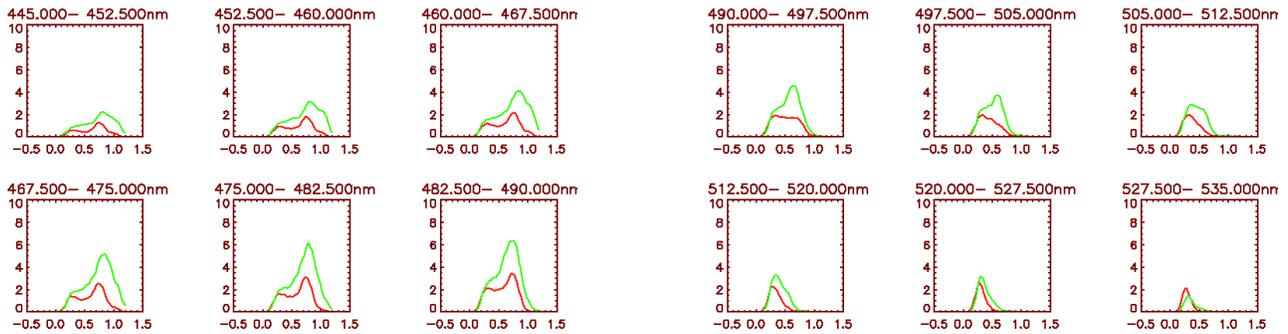

**Fig 2.** Plotted is Raman backscatter power vs. time in contiguous wavelength bins. This is wavelength partitioning of the witness beam's SRBS when a pump and probe beam pair cross to resonantly generate an EPW (Red) and when the blue pump beam is turned off (Green). We find factors of 3 reduction in SRBS in the presence of an externally generated EPW in a wavelength bin and time window which is consistent with the peak of the plasma density profile being at the right density where the EPW could have been driven by crossing blue and green beams, as inferred from LASNEX simulations. Green beam energy was 100J (probe) and the blue beams (pump and witness) were at 500J.

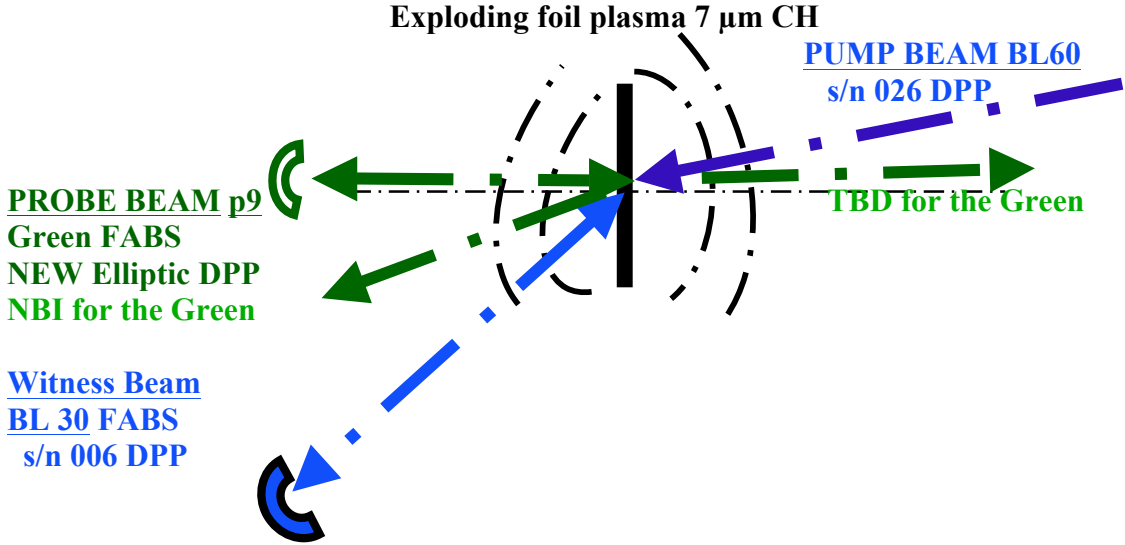

**Figure 3.** This is a schematic diagram of the experimental configuration of the three interaction beams used (crossing pump-BL60, probe-green, and witness-BL30 beams) and the optical diagnostics fielded. Full aperture backscatter stations (FABS) for SRS and SBS were used on the blue witness beam BL30 and the green probe beam. Near backscatter imaging (NBI) on BL30 and the green beam as well as a transmission beam diagnostic (TBD) for the green beam alone were also deployed. All three interaction beams had distributed phase plates or DPPs. BL60 is the pump beam which amplifies the green probe and resonantly excites an EPW via optical mixing with the green. Its backscatter is not measured due to space and other material resource constraints in the target chamber on Omega at LLE.

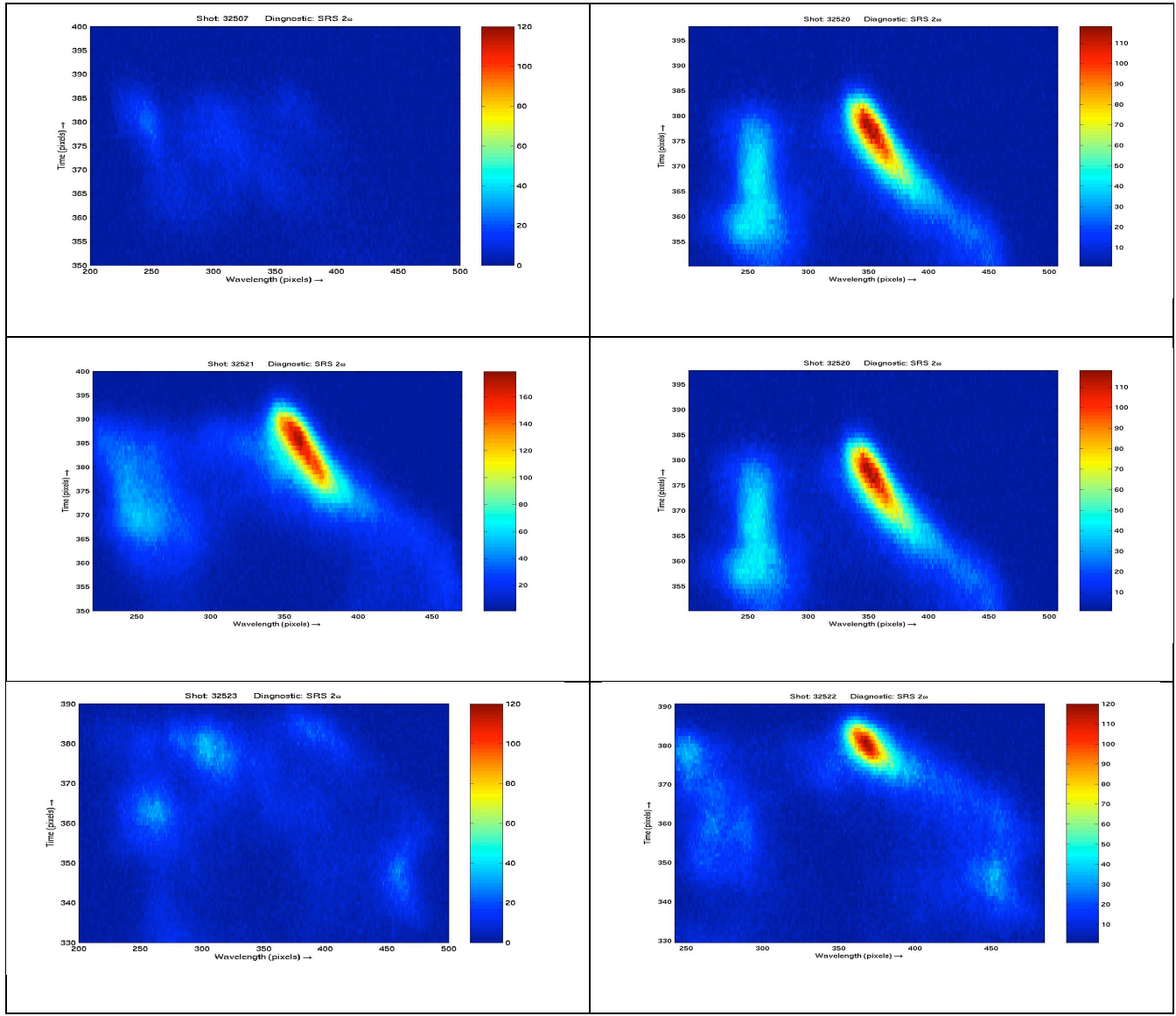

**Figure 4.** Green beam SRS backscattering comparisons. Optical mixing amplified green beam Raman backscattering (SRBS) is observed when the incident green beam energy is 94J. At this energy, green beam SRBS is very weak. However, when a large amplitude blue pump beam is also present crossing the green from the opposite side, it amplifies the green beam to such as extent (around a factor of two) that now the green beam's SRS backscatter is fifteen times higher in peak magnitude. The second row compares cases with two blue pump beam energies and shows similar results with enhanced filamentation signature SRS coming from very low density plasma when the blue pump beam is on. The third row is similar to the first except the hydrodynamics of the target was changed due to two heater beams being dropped in order to accommodate the green beam TBD diagnostic. This made the density of the plasma 25% higher at corresponding times during the pulse as well as the temperature lower, which promoted larger levels of filamentation. Despite all that, remarkably similar backscatter enhancement is seen in the presence of the Blue crossing beam. The hydro was calculated using LASNEX. Row 1: $E_G$ = 94 J, $E_{BL60}$= 0 and 489 J. Row 2: $E_G$ = 94 J, $E_{BL60}$=374 J and 489 J. Row 3: $E_G$ = 94 J, $E_{BL60}$=0 and 485 J.